# Modeling of protein-peptide interactions using the CABS-dock web server for binding site search and flexible docking

Authors: Maciej Blaszczyk, Mateusz Kurcinski, Maksim Kouza, Lukasz Wieteska, Aleksander Debinski, Andrzej Kolinski, Sebastian Kmiecik*

Authors Affiliation: Department of Chemistry, University of Warsaw, Pasteura 1, 02-093 Warsaw, Poland

* corresponding author, email: sekmi@chem.uw.edu.pl, address: Department of Chemistry, University of Warsaw, Pasteura 1, 02-093 Warsaw, tel: +48 22 822 02 11 ext. 310; fax: +48 22 822 02 11 ext. 320

**ABSTRACT**

Protein-peptide interactions play essential functional roles in living organisms and their structural characterization is a hot subject of current experimental and theoretical research. Computational modeling of the structure of protein-peptide interactions is usually divided into two stages: prediction of the binding site at a protein receptor surface, and then docking (and modeling) the peptide structure into the known binding site. This paper presents a comprehensive CABS-dock method for the simultaneous search of binding sites and flexible protein-peptide docking, available as a user's friendly web server. We present example CABS-dock results obtained in the default CABS-dock mode and using its advanced options that enable the user to increase the range of flexibility for chosen receptor fragments or to exclude user-selected binding modes from docking search. Furthermore, we demonstrate a strategy to improve CABS-dock performance by assessing the quality of models with classical molecular dynamics. Finally, we discuss the promising extensions and applications of the CABS-dock method and provide a tutorial appendix for the convenient analysis and visualization of CABS-dock results. The CABS-dock web server is freely available at http://biocomp.chem.uw.edu.pl/CABSdock/.

# 1. INTRODUCTION

Despite the significant progress in experimental and theoretical studies of protein-peptide interactions, the understanding of their role in the cellular machinery remains quite limited. Over the years it has become clear that understanding a particular protein function as a combination of its functional domains is not complete [1]. It has been shown that in higher eukaryotes up to 40% of protein-protein interactions (PPIs) are mediated by peptides [2]. Peptides responsible for PPIs are not necessarily independent molecules, but more often appear as disordered regions within proteins (at termini, between domains or flexible loops) that can act as a separate peptide molecule. The view that proteins can be understood through their discrete segments has already provided important insights into protein function [1]. Especially, protein-peptide interactions can be found in intracellular signaling pathways, cell localization, immune response, and protein degradation. Their new functional roles are constantly being discovered [3]. Importantly, many of these interactions are implicated in human diseases such as cancer or autoimmune disease [4-7]. Therefore, structure-based studies directed toward the design of completely new or modified receptor-interacting peptides have become a hot spot of current biomedical research.

In comparison to PPIs, protein-peptide interactions are more transient and interaction affinity is significantly weaker. Together with the high conformational flexibility of peptides, these factors make structural characterization of protein-peptide complexes really challenging. Therefore, there is an urgent need for the development of complementary computational approaches, such as effective molecular docking [8]. Assuming that the structure of a protein receptor has been solved experimentally or modeled with good accuracy, the modeling protocol for searching new protein-peptide interactions usually has two or three major steps:

(1) The first step involves identification of the binding site on the protein surface. This goal can be accomplished by bioinformatics methods using data from already known protein and protein-peptide structures or simply protein sequences [9-12]. They mostly aim at creating a library of sequence, structure or surface landscape motifs that could be universally detected in unknown proteins [13-15]. It should be noted that due to its simplicity, information only about sequence patterns that occur within binding sites is not sufficient for accurate binding site prediction and could result in a high ratio of false-positives [1].

(2) Second, the peptide is docked to a known binding site using local docking techniques, such as adapted Molecular Dynamics [16], Rosetta FlexPepDock [17, 18], HADDOCK [19, 20] or PepCrawler [21] methods (see reviews [2, 8]).

(3) Third, those methods for local peptide docking may also serve in the final modeling step: high resolution refinement of initially generated peptide poses.

The first two steps of modeling protein-peptide interactions can also be achieved using techniques for the combined search of binding sites and peptide poses [13, 22, 23]. Usually, these methods allow the identification of a binding site, although the quality of resulting peptide models is often unsatisfactory [2].

Recently, we have developed a CABS-dock method and a web server for the simultaneous prediction of binding sites and protein-peptide docking [24]. The CABS-dock simulation engine, based on the coarse-grained CABS model, enables efficient docking search of fully flexible peptides over the entire surface of flexible proteins in a reasonable time, typically 1 to 8 CPU hours (which is thousand-fold shorter than analogical simulations using rapid molecular dynamics adapted to peptide docking [22]). CABS-dock has been extensively tested over the largest benchmark set of non-redundant protein-peptide interactions available to date (including docking to bound and unbound receptor forms). For over 80% of bound and unbound dataset cases, we obtained high or medium accuracy models (expected to be of sufficient accuracy for high resolution refinement) [24].

In comparison to other protein-peptide docking tools (listed above), the CABS-dock offers the following major advantages: (1) the method does not require knowledge of the binding site nor any information about the peptide conformation, (2) during docking peptide conformation is allowed to be fully flexible, and (3) it is possible to simulate significant conformational changes of the protein receptor structure (see section 3.2.1). These advantages become even more apparent in comparison to general purpose protein-ligand docking tools, which are usually less efficient in sampling conformational changes than methods dedicated to protein-peptide docking. The possible CABS-dock disadvantages include: (1) lack of option to guide the docking with the knowledge of the binding site (this will be available in the next CABS-dock update planned in 2015), however, it is possible to exclude some receptor areas from the docking search, and thereby to enforce more effective search in a closer neighborhood of the potential binding site (see section 3.2.2); (2) a small set of 10 best scored models may not show the high accuracy models (that may be present in the large set of CABS-dock predicted models), however, this is also the case for the other docking methods (scoring problem is discussed in section 3.4).

In this work, we evaluate CABS-dock performance and focus on particular examples of protein-peptide docking. The examples discussed illustrate CABS-dock performance using the default server mode as well as its advanced options. We also address the possibility of improving CABS-dock performance using an external scoring method over a large set of CABS-dock generated models. This can be achieved by a two-step procedure involving: (1) reconstruction and local optimization of CABS-dock models, followed by their (2) scoring using short simulations by all-atom molecular dynamics with explicit solvent. An Appendix is also provided with this paper where we provide a tutorial for the display and analysis of CABS-dock models and trajectories using VMD [25], a molecular graphics program.

## 2. METHOD

The CABS-dock method [24] is based on the CABS model (described in detail in ref. [26]) that was originally designed for the structure prediction of globular proteins and simulation of protein dynamics. CABS comes from the letters of pseudo-atoms used to represent a single protein residue: carbon alpha (CA), carbon beta (B) and side chain (S). An additional pseudo-atom, defined in the geometrical center of the virtual CA-CA bond, is used to define the main chain hydrogen bonds. To speed up calculations, the coordinates of the CA atoms are restricted to the beads of a dense cubic lattice with lattice spacing arbitrary set to 0.61Å. The remaining pseudo-atoms are located off the lattice and follow the movement of the main chain. The force field is based on knowledge-based statistical potentials derived from structural regularities seen in known protein structures. Sampling is controlled by the asymmetric Metropolis criterion. Additionally, CABS uses the Replica Exchange protocol for better sampling coverage of the energy landscape.

The CABS model was initially used for protein structure prediction and it performed exceptionally well in CASP6 (Critical Assessment of protein Structure Prediction, a community-wide blind test of structure prediction approaches). Using the CABS-based approach, the Kolinski-Bujnicki group scored best or second best, depending on the evaluation method [27, 28]. CABS-based protocols for the *de novo* and consensus prediction of protein structure were made freely available to the academic community on an automated web server [29]. The CABS model has also been successfully used to simulate the dynamics of denatured protein states [30], protein folding mechanisms [31-36], the flexibility of globular proteins [37-39] and its influence on protein aggregation [40]. Finally, the CABS model has been optimized for the investigation of protein interactions and prediction of structures of protein complexes. It has been used to build a model of human telomerase [41], protein-peptide and protein-protein docking [42-45] and to investigate the mechanism of simultaneous folding and binding of an intrinsically disordered peptide [46].

The pipeline of the CABS-dock server is a multistage protocol that consists of multiple programs and associating scripts, with the CABS model (version dedicated to handle multimeric protein chains) at its center. As shown in Figure 1, the whole procedure consists of four main stages 1) flexible docking by the CABS algorithm; 2) initial filtering of probable solutions from all generated models; 3) further selection of representative models by the clustering protocol and 4) reconstruction to all-atom representation and local optimization of final models. The method is fully automated. As an input it requires only the 3D structure of the receptor and the peptide sequence. On output the server returns 10 top scored models of the protein-peptide complex. Each step of the procedure is briefly described in the following paragraphs.

**Figure 1. CABS-dock pipeline.** Sets of models generated with the pipeline are available for download from the "Docking prediction results" tab ("Download all files" button).

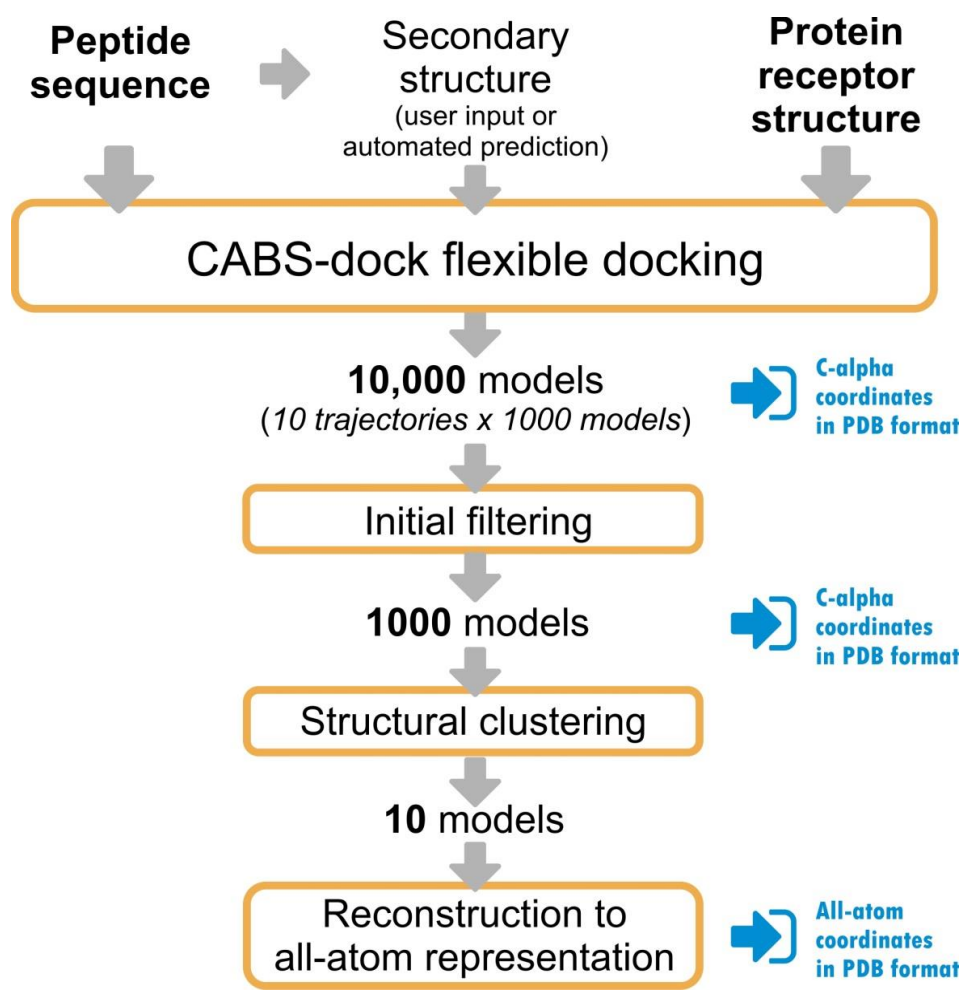

## 2.1 Flexible docking with the CABS model

The CABS-dock method requires two inputs: (1) amino acid sequence of the peptide, and (2) 3D structure of the protein receptor (the obligatory and non-obligatory input recommendations are listed in [24]). In the first CABS-dock modeling step, 10 copies of the protein-peptide system are generated as starting models for the Replica Exchange Monte Carlo sampling method. Each starting copy contains a random peptide structure that is placed in a random position within 20 Å from the input receptor structure (see Figure 2, left). During simulation of coupled peptide binding and folding, the CABS-dock protocol allows full flexibility of the peptide and small fluctuations of the receptor backbone. The simulation result is the set of 10,000 models (visualized in Figure 2, middle) in the C-alpha representation, collected in 10 trajectories. Each trajectory counts 1000 models and shows system evolution for each replica.

**Figure 2. Molecular visualization of the basic stages of the CABS-dock protocol.** Three stages are shown: 10 starting random peptide conformations (left), 10,000 models (middle), 10 final models (right). The protein receptor is showed as gray surface, peptides in cyan. The picture shows the human androgen receptor as an example (PDB ID: 2AM9).

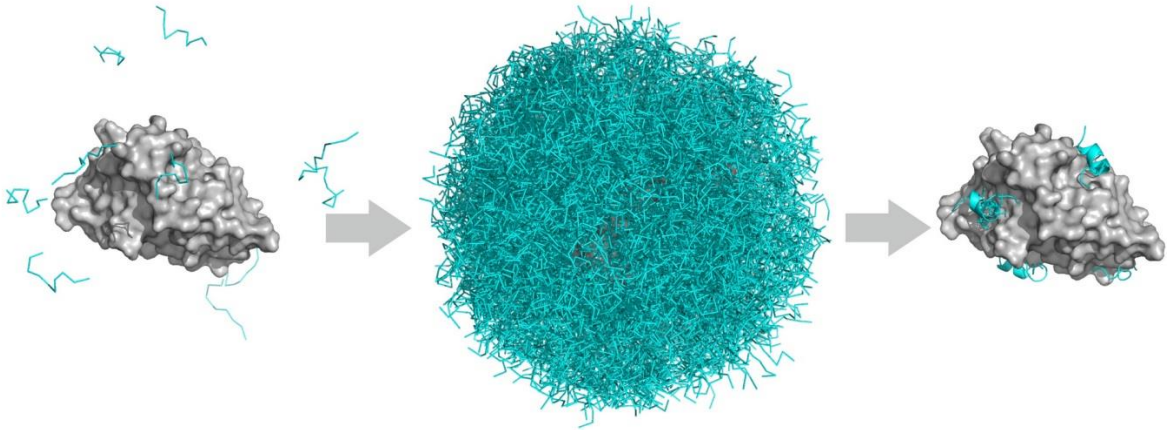

## 2.2 Initial filtering

From the 10,000 models generated during the simulation up to 1000 are selected for further steps in the following procedure: 1) all unbound states (where interaction energy between the peptide and the receptor is zero) are rejected 2) from the remaining models up to 100 from each of the 10 trajectories are picked by the lowest interaction energy. This procedure usually retains the best of the generated models in the filtered set (see example in Figure 3).

**Figure 3. Relationship between energy and ligand-RMSD values in an example CABS-dock simulation** (example for protein PDB ID: 2AM9). The top panels show energy values (from the CABS-dock force-field) for the entire complex structure, while the bottom panels for protein-peptide interaction only. The left panels present the data for all models generated in the CABS-dock simulation (10,000), while the right panels for 1000 models selected in the initial filtering procedure (see the pipeline, Figure 1). Ligand-RMSD is the root mean square deviation calculated on the peptides after superposition of the receptor molecules. Note that for each prediction case a similar plot can be easily created by the user (see Appendix for instructions).

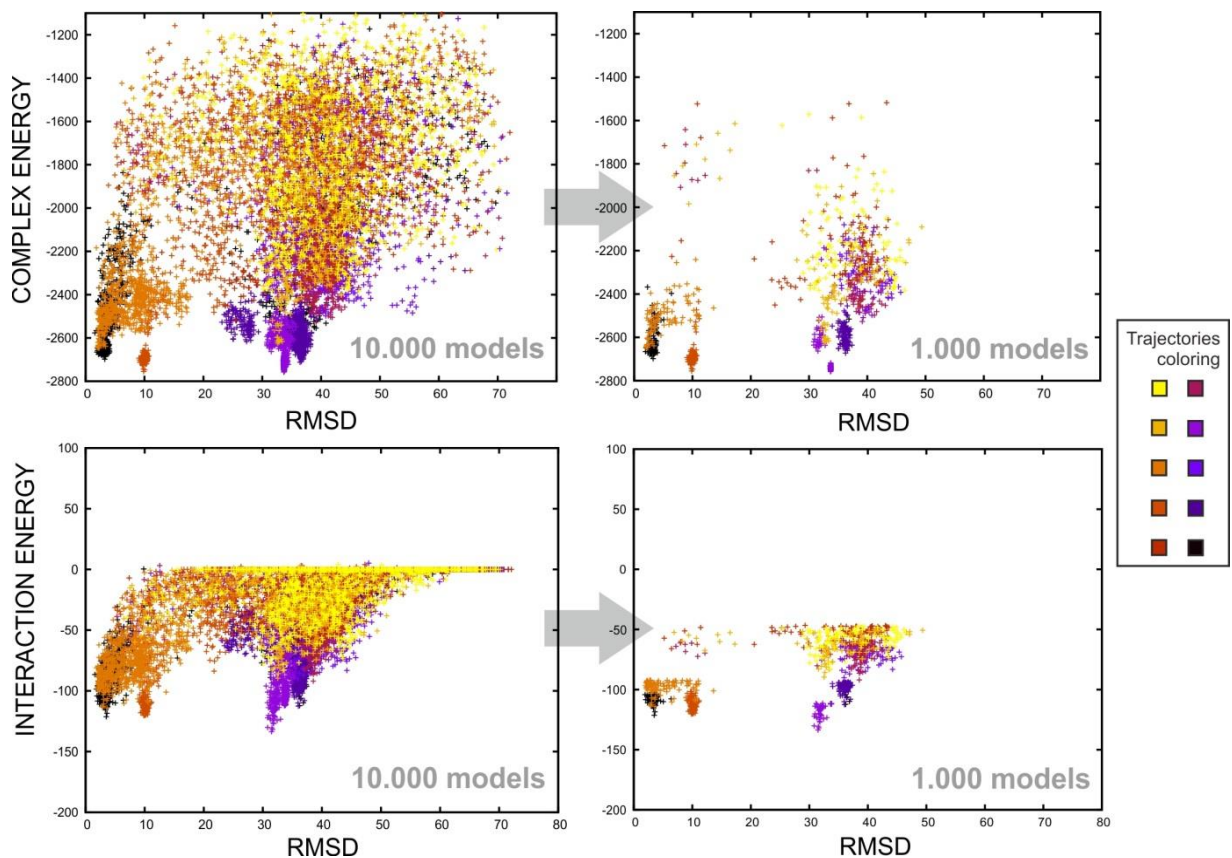

### 2.3 Clustering

The 1000 filtered models are subsequently grouped into clusters in the k-medoid clustering procedure. The clustering is run 100 times with different initial seeds and k=10. Consensus medoids are selected as the final models. The density of the clusters (defined as an average difference between cluster elements divided by the number of elements) is used to finally rank the models. Ligand-RMSD (root mean square deviation calculated on the peptides after superposition of receptor molecules) is used as the differentiation measure between cluster elements.

### 2.4 Reconstruction and local optimization of the final models

The last step of the CABS-dock protocol (also discussed in section 3.3) is the all-atom reconstruction of the final models using the MODELLER [47] program. 10 medoids serve as templates for the reconstruction and optimization procedure in the DOPE potential [48]. As a result, energy minimized all-atom representations of the final models are obtained.

### 3. FEATURES AND APPLICATIONS

### 3.1 Docking without prior knowledge of the binding site

The recent publication on the CABS-dock server [24] describes CABS-dock performance (with default server settings) over a large dataset of peptide-protein complexes, including docking to bound and unbound (when available) forms of protein receptors. For over 80% of bound and unbound cases we obtained high or medium accuracy models (see Figure 4).

**Figure 4. CABS-dock performance summary for 103 bound and 68 unbound benchmark cases.** The quality assessment criteria are based on ligand-RMSD (root-mean-square deviation) between the predicted model and the experimental peptide structure after superimposition of the receptor molecules (high accuracy: ligand-RMSD<3 Å; medium accuracy: 3 Å ≤ ligand-RMSD ≤ 5.5 Å; low accuracy: ligand-RMSD > 5.5 Å). The percentages shown refer to the best quality models found in the set of all 10,000 generated models (**all**) and in the set of 10 top-scored models (**top 10**). More details can be found in our recent publication [24] and in online materials at http://biocomp.chem.uw.edu.pl/CABSdock/benchmark.

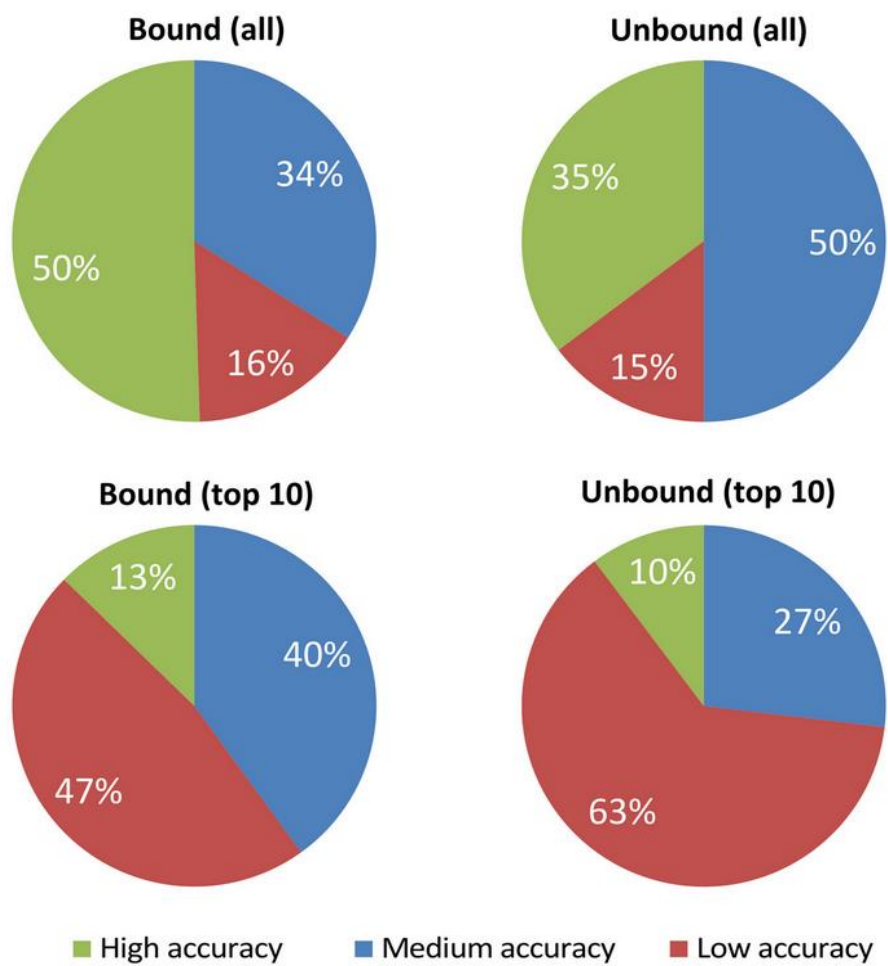

In the next subsections, detailed analysis of modeling cases selected from the benchmark dataset [24] is presented. All CABS-dock predictions for the entire benchmark dataset are available for download from http://biocomp.chem.uw.edu.pl/CABSdock/benchmark

### 3.1.1 An example when the accurate model is top ranked

The example described below has been obtained with the default CABS-dock server settings using the following input data:

- **Peptide sequence:** SSRFESLFAG
- **Peptide secondary structure:** CHHHHHHHHC

- **Receptor input structure:** PDB ID: 2AM9, crystal structure of human androgen receptor in the unbound form

Reference structure used for the calculation of ligand-RMSD values to the experimentally determined peptide-bound structure:

- **Peptide-receptor complex structure:** PDB ID: 1T7R, crystal structure of human androgen receptor in complex with the peptide.

In 10 final models, peptides were docked to five different binding sites (see Figure 5). In two of the models (including the first, top-ranked model) the peptide was bound in the correct binding site. As shown in Table 1, the one top ranked (representative of the most dense cluster) is the most accurate. More details about cluster content can be displayed under the "Clustering details" tab (see Figure 6).

**Figure 5. CABS-dock modeling with default settings – an example when the accurate model is top ranked.** The figure shows an experimental protein peptide-bound form (the receptor is colored in gray, peptide in magenta, PDB ID: 1T7R) together with CABS-dock-predicted peptide poses (colored in cyan). Top 10 CABS-dock models are presented, which were docked in five potential binding sites (one of the peptide models is not visible because it is docked at the opposite receptor surface). In the native binding site (marked by the rectangle), two models were docked. One of these models (which is the top-ranked first model, representative of the top ranked cluster, see Table 1) is presented on the right with the experimental peptide structure. Ligand-RMSD between the first model and the experimental peptide structure is 2.22 Å.

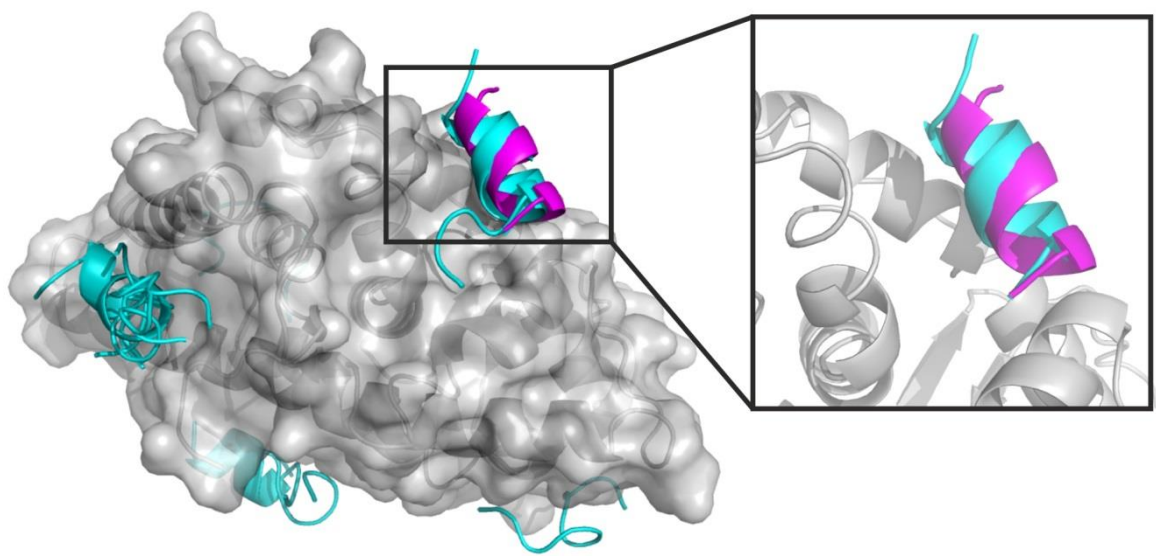

**Table 1. Example details of structural clustering.** Clusters are ranked according to cluster density. In the present case the most dense cluster is the most numerous one (226 models out of 1000) and the most similar to the experimental model. The table shows data for the prediction case described in Figure 5. Ligand-RMSD values are also presented (root mean square deviation calculated on the peptides after superposition of receptor molecules).

| Cluster name | Cluster density | Average cluster RMSD (Å) | Maximum RMSD within the cluster (Å) | Number of cluster elements | Ligand-RMSD between the medoid and experimental form (Å) |
|---|---|---|---|---|---|
| **Cluster 1** | 33.68 | 6.70 | 34.07 | 226 | 2.22 |
| **Cluster 2** | 26.33 | 2.31 | 7.23 | 61 | 33.63 |
| **Cluster 3** | 22.64 | 6.05 | 25.11 | 137 | 9.81 |
| **Cluster 4** | 11.93 | 8.04 | 33.71 | 96 | 34.46 |
| **Cluster 5** | 11.90 | 8.06 | 24.09 | 96 | 39.20 |
| **Cluster 6** | 11.83 | 6.93 | 29.92 | 82 | 31.63 |
| **Cluster 7** | 11.29 | 6.81 | 16.26 | 77 | 37.91 |
| **Cluster 8** | 10.98 | 8.37 | 28.01 | 92 | 38.85 |
| **Cluster 9** | 9.02 | 9.42 | 34.85 | 85 | 33.97 |
| **Cluster 10** | 6.39 | 7.50 | 14.25 | 48 | 32.17 |

**Figure 6. Screenshot of the "Clustering details" tab of the CABS-dock web server.** An interactive chart shows the composition of clusters of models vs. their trajectory affiliation. By clicking on the points representing models in the chart, particular models can be viewed in 3D and downloaded in the pdb format. In this case model number 828 from the first trajectory is shown, which is assigned to the first, the most dense cluster (see also Table 1). This model is the best among the 1000 top scored models (ligand-RMSD to the experimental peptide structure: 1.93 Å).

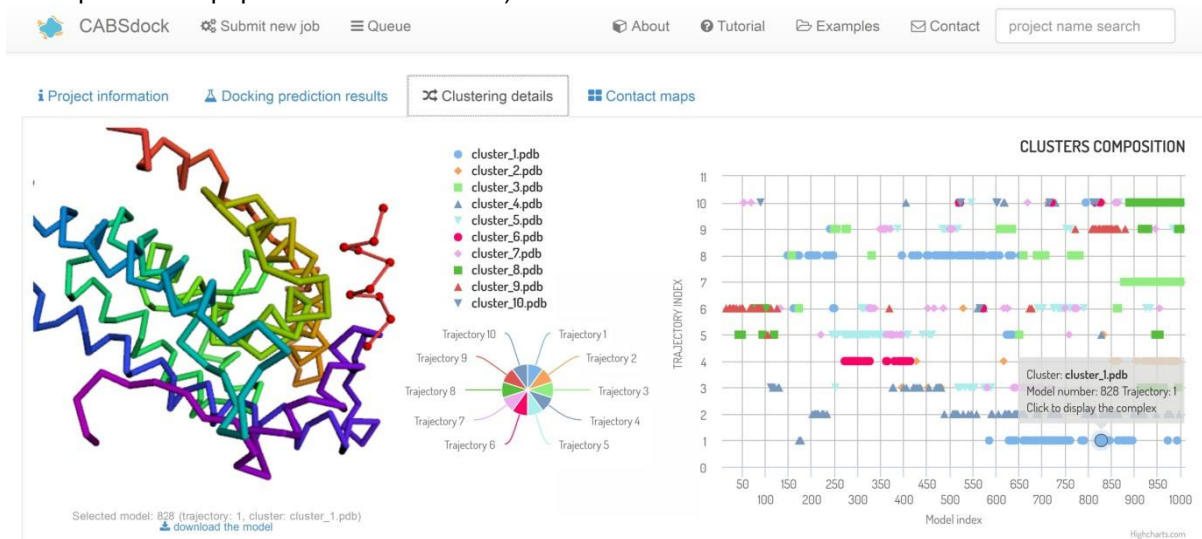

### 3.1.2 An example when a medium accuracy model is top ranked and a high accuracy model exists in the trajectory

Example 3.1.1 shows the most favorable situation when a high-accuracy model was identified as the first model in the final top 10. However, due to the high complexity of the problem, high-accuracy models may be missing among the top 10 models, but they may exist in simulation trajectories (in a set of 10,000 models).

The example described below was obtained with the default CABS-dock server settings using the following input data:

- **Peptide sequence:** RRNLKGLNLNLH
- **Peptide secondary structure:** CCCCCCCCCCCC
- **Receptor input structure:** PDB ID: 2B9F, crystal structure of mitogen-activated protein kinase FUS3 in the unbound form

Reference structure used for the calculation of ligand-RMSD values to the experimentally determined peptide-bound structure:

- **Peptide-receptor complex structure:** PDB ID: 2B9H, crystal structure of mitogen-activated protein kinase FUS3 in the peptide-bound form.

In the 10 final models, peptides were docked to five different binding sites; however, only one model (representative of the second cluster) was close to the experimental peptide binding site with a ligand-RMSD value of 4.16 Å to the reference peptide structure in the 2B9H complex (see Figure 7 and Table 2). A more accurate model can be found in the set of all 10,000 models generated by CABS-dock with a ligand-RMSD value of 2.33 Å. In such modeling cases, external scoring methods may be useful for fishing out the best accuracy models from the large pool of generated models (see section 3.4).

**Figure 7. CABS-dock modeling with default settings - an example when a medium accuracy model is top ranked and a high accuracy model exists in the trajectory.** The figure shows an experimental protein peptide-bound form (receptor is colored in gray, peptide in magenta, PDB ID: 2B9H), together with CABS-dock-predicted peptide poses (colored in cyan) and the most accurate prediction from the entire trajectory (in green). 10 top ranked models have peptides docked in five different areas. One of these areas is the native binding site (marked by the rectangle) with a medium-accuracy model (model 2, ligand-RMSD to the experimental structure: 4.16 Å). On the right, the native binding site is shown with the most accurate prediction from all simulation data (ligand-RMSD between the predicted and experimental peptide structure: 2.33 Å).

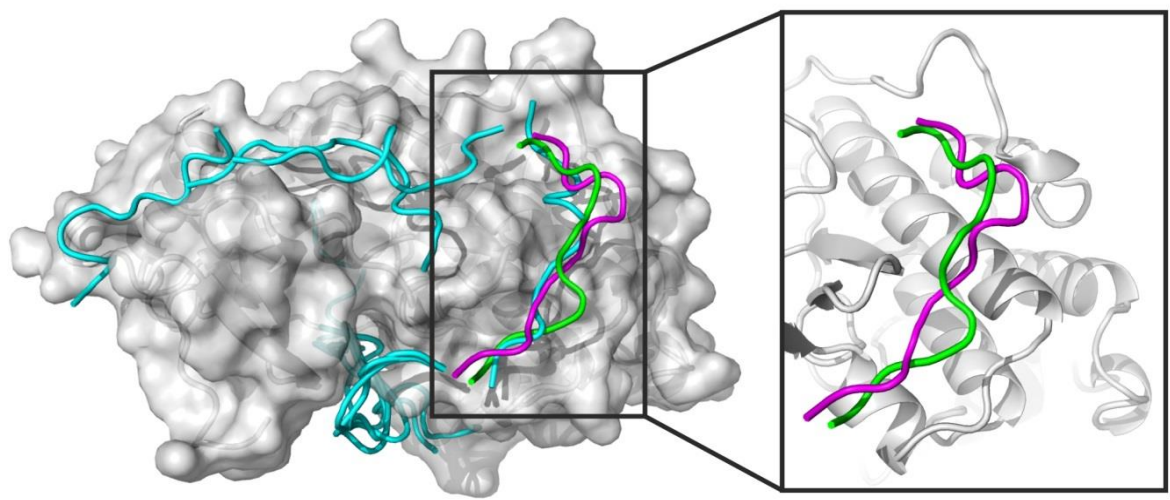

**Table 2. Example details of structural clustering (as displayed in the CABS-dock server, column containing ligand-RMSD to the experimental form added).** Clusters are ranked according to cluster density. The cluster whose medoid is the most similar to the experimental model is the most numerous one, but ranked as second (according to cluster density). The table shows data for the prediction case described in Figure 7.

| Cluster name | Cluster density | Average cluster RMSD (Å) | Maximum RMSD within the cluster (Å) | Number of cluster elements | Ligand-RMSD between the medoid and experimental form (Å) |
|---|---|---|---|---|---|
| **Cluster 1** | 31.96 | 1.15 | 2.04 | 37 | 23.87 |
| **Cluster 2** | 23.71 | 6.45 | 39.57 | 153 | 4.16 |
| **Cluster 3** | 19.10 | 7.28 | 31.51 | 139 | 21.25 |
| **Cluster 4** | 15.79 | 8.87 | 22.47 | 140 | 31.77 |
| **Cluster 5** | 13.01 | 10.99 | 21.00 | 143 | 19.96 |
| **Cluster 6** | 12.02 | 10.48 | 29.26 | 126 | 39.62 |
| **Cluster 7** | 10.20 | 12.54 | 55.41 | 128 | 37.72 |
| **Cluster 8** | 9.34 | 10.28 | 43.88 | 96 | 28.23 |
| **Cluster 9** | 7.70 | 1.039 | 2.82 | 8 | 23.32 |
| **Cluster 10** | 5.74 | 5.22 | 17.13 | 30 | 22.99 |

## 3.2 CABS-dock modeling with advanced options

Together with its basic functionality, CABS-dock enables advanced modification of docking simulation settings. This is done using two advanced options for: (1) increasing the level of flexibility for selected receptor fragments, (2) excluding user-selected binding modes from docking search.

### 3.2.1 Increasing the flexibility of receptor fragments

The advanced option for increasing flexibility for selected receptor residues is available from the main page by checking the "Mark flexible regions" option.

For each selected residue, the user may choose from two preset settings: moderate or full flexibility. Technically this is achieved by changing the default distance restrains (used to keep the receptor structure near to the input conformation). The assignment of moderate flexibility decreases the strength of restrains, while full flexibility assigned removes all the restraints imposed on the selected residue.

Below, we describe a practical example of using the "Mark flexible regions" option with the following input data:

- **Peptide sequence:** HPQFEK
- **Peptide secondary structure:** CHHHCC
- **Receptor structure:** PDB ID: 2RTM, crystal structure of biotin binding protein in the unbound form

Reference structure used for the calculation of ligand-RMSD values to the experimentally determined peptide-bound structure:

- **Peptide-receptor complex structure:** PDB ID: 1KL3, crystal structure of biotin binding protein in the peptide-bound form.

According to the experimental studies [49] the unbound form of biotin binding protein has a flexible loop close to the binding site. Using the CABS-dock "Mark flexible regions" option, we selected 10 residues (from $45^{th}$ to $54^{th}$) forming the flexible loop and assigned the "fully flexible" setting to those residues.

As shown in Figure 8, the initial position of the loop in the unbound form would prevent correct binding of the peptide. Assigning full flexibility to the loop residues enabled us to uncover the binding site and to obtain a high accuracy model (ligand-RMSD of the first top-ranked model to the experimental structure: 2.03 Å).

**Figure 8. CABS-dock modeling with full flexibility of a protein loop region close to the binding site.**
(a) Comparison of the experimental protein structure in the peptide-unbound form (colored in green; CABS-dock input structure, PDB ID: 2RTM) with a peptide-bound experimental complex (in magenta, PDB ID: 1KL3) and a CABS-dock-predicted complex (in pale cyan). Peptide backbones are presented as thick lines, while loop backbones as thin lines. Ligand-RMSD between the predicted and experimental peptide structure is 2.03 Å. (b) Loop region flexibility in CABS-dock modeling. Protein structures from CABS-dock predicted models (in pale green) are compared with the unbound protein form (in green). The flexible loop region (designated to be fully flexible during docking) is marked in red (residues 45 to 54, constituting a region 10 residues in length).

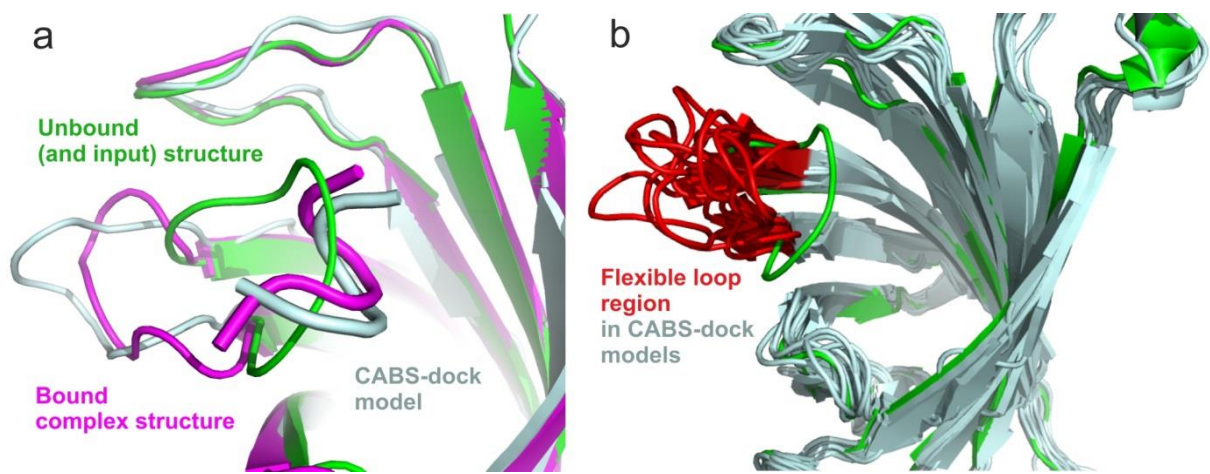

### 3.2.2 Excluding binding modes from docking search

In the default mode, CABS-dock allows peptides to explore the entire receptor surface. However, in certain modeling cases it is known that some parts of the protein are not accessible (for example due to binding to other proteins) and therefore it could be useful to exclude these regions from the search procedure. Such a feature is available by selecting the "Mark unlikely to bind region" option on the main page. This option works in two ways:

1) by selecting the residues to be excluded (available from the main page by checking the "Mark unlikely to bind regions" option)
2) by re-submitting a previously run job (a Resubmit button is available in the "Project information" tab) and selecting resulting models (binding modes) to be excluded from future results. Thus, in practice, the excluding option can be also used to force the CABS-dock algorithm to search for additional binding sites not found in the previous runs.

Below, we describe a practical example of the excluding option by re-submitting a previously run job.

- **Peptide sequence:** PQQATDD
- **Peptide secondary structure:** CEECCCC
- **Input structure:** PDB ID: 1CZY:C, tumor necrosis factor receptor associated protein 2 in the unbound form

For the calculation of ligand-RMSD values to the experimental peptide binding pose, we used:

- **Reference structure:** PDB ID: 1CZY:CD, tumor necrosis factor receptor associated protein 2 in the peptide-bound form

The first simulation run resulted in 10 top-ranked models having peptides bound mostly in a single area far from the native binding site (see Figure 9). Excluding these peptide poses (by re-submitting a previously run job) resulted in new predictions among which the first top-ranked model was consistent with experimental structure.

**Figure 9. CABS-dock modeling with excluding binding modes from previous prediction runs.** The figure shows an experimental receptor peptide-bound form (PDB ID: 1CZY; receptor shown as surface, peptide as a magenta line) together with CABS-dock peptide poses predicted in the first and re-submitted run (performed by excluding peptide poses from the first run). Predicted peptide poses are shown in red (8 poses obtained in the first run excluded from the re-submitted job), green (2 poses from the first run not excluded from the re-submitted job) and in cyan (result of the re-submitted job). In the native binding site (marked by the rectangle), two models were docked. One of these models (which is the top-ranked first model, representative of the top ranked cluster) is shown on the right with the experimental peptide structure. Ligand-RMSD between the first model and the experimental peptide structure: 2.89 Å.

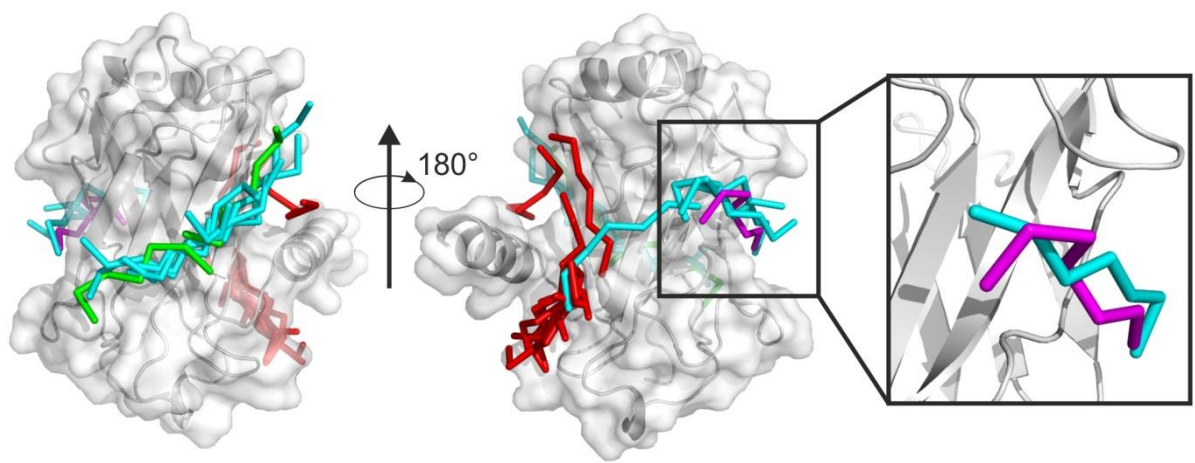

## 3.3 All-atom reconstruction and local optimization
### 3.3.1 Accessing CABS-dock models in coarse-grained representation

The main output of the CABS-dock server is a set of 10 top scored models in all-atom representation. These models are representatives of a broad set of 10,000 initial models and are selected through stepwise filtering and clustering procedures (see Figure 1 and Methods section). From the "Docking predictions results" tab, a compressed archive containing the complete set of CABS-dock generated models in coarse-grained representation can be downloaded. The archive consists of the following sets of models (see the Appendix for a detailed description of the archive file):

- model_*.pdb –10 final models, numbered from 1 to 10 (in the PDB file format and all-atom representation)
- cluster_*.pdb – cluster models (groups of models that have been classified in structural clustering to particular clusters), numbered from 1 to 10 (in the PDB file format and C-alpha representation). Cluster numbering corresponds to model numbering, i.e. model_7.pdb is a representative model of models grouped in the seventh cluster (ranked as seventh) (cluster_7.pdb).
- trajectory_*.pdb – complete set of 10 trajectories, numbered from 1 to 10 (in the PDB file format and C-alpha representation). Each trajectory contains 1000 models.
- top1000.pdb – top 1000 models (selected for further clustering and analysis) from the 10 trajectories (in the PDB file format and C-alpha representation)

All the sets of models in coarse-grained representation can be reconstructed to physically-sound all-atom models, as we demonstrated in [34, 50]. Those reconstructed models can be later used in applications that require all-atom details (e.g. studies of energy properties using all-atom modeling tools, or as an input for all-atom analysis and visualization).

### 3.3.2 Strategies for reconstruction and local optimization

Various strategies can be employed for the reconstruction of protein models from C-alpha trace to all-atom models. Successful approaches in the application to CABS-generated models are presented in [34, 50]. The reconstruction of atomic details is typically a two-stage process, where at first only the backbone atoms are reconstructed [51]. In the next step the remaining atoms of the side chains are added to the backbone and a complete model is subsequently optimized to remove structural inaccuracies, such as improper bond lengths and angles, or steric clashes.

There are several methods available for protein backbone reconstruction from C-alpha coordinates using fragments taken from known protein structures [51-58]. An important attribute of the method chosen for backbone rebuilding should be insensitivity for small (unphysical) local distortions of C-alpha distances present in CABS generated models [34, 50]. Such distortions are for example well-handled using the approach of Claessens et al. [53] or the ModRefiner tool [59] (which unfortunately doesn't handle multimeric protein chains). For the second rebuilding step, side chain reconstruction, the SCWRL program can be recommended [60].

The reconstruction of models should be followed by local structural optimization [61]. Ideally, a fully reconstructed model should undergo a short MD simulation with explicit solvent. However, for practical applications such an approach may be computationally too demanding, especially when thousands or more models need to be processed. Therefore, efficient algorithms handling the entire reconstruction and optimization process have also been proposed [54, 59, 62].

Rebuilding models after protein-peptide docking raises another challenge: dealing with multiple protein chains. However, this is rather a technical than conceptual problem as most of the above-listed tools are not capable of optimizing the structure of a multi-chain protein complex. In the CABS-dock server, we have implemented a procedure based on the Modeller package [47] for combined reconstruction and optimization. The optimization step is done by the minimization of the DOPE (Discreet Optimized Protein Energy) statistical potential [48]. The script for the Modeller reconstruction and optimization procedure is available in the CABS-dock online tutorial at http://biocomp.chem.uw.edu.pl/CABSdock/tutorial.

### 3.3.3 Database of large sets of CABS-dock models in an all-atom representation

Large sets of CABS-dock generated models can be used for the all-atom scoring of protein-peptide interactions or for the development of new scoring potentials. For such exercises, we have rebuilt to all-atom representation sets of top-scored models (comprising 100 or 1000 models) for the entire benchmark set of protein-peptide complexes (103 bound cases and 68 unbound cases that have been used for validation of CABS-dock performance [24]). The benchmark sets of all-atom models are available for download at http://biocomp.chem.uw.edu.pl/CABSdock/benchmark.

### 3.4 Scoring of predicted models
### 3.4.1 Results of scoring exercises using all-atom MD with explicit solvent

One of the main problems in modeling protein interactions is the scoring problem [8, 63, 64]. In general terms, it is usually a problem of selection of predicted models that are closest to the native (real) complex out of a large set of diverse models. There are three kinds of scoring functions: physics-based [65], empirical [66], and knowledge-based [67]. The former two calculate binding energy as a sum of individual energy terms. Usually, in the physics-based approach, summation of Lennard-Jones and Coulombic potential functions from popular force-fields is used. Empirical scoring functions involve a sum of weighted uncorrelated terms, trained to reproduce the known experimentally determined protein-peptide binding affinities. Knowledge-based scoring functions are based on statistical observations of intermolecular interfaces found in known protein complexes. In many physics-based scoring functions [68-70], water is not treated explicitly, and the solvent effect is either neglected or treated as a continuum dielectric medium. The golden standard to account for the solvent effect is to use explicit solvent molecules in an all-atom MD technique. While using this technique involves a sampling problem (due to the large system size and consequent timescale limitations), it can be implemented as a short (computationally inexpensive) simulation procedure in combination with coarse-grained simulations [36].

For five cases of unbound docking (from the benchmark results [24]), we examined the possibility of scoring 100 models (top-scored by CABS-dock) using all-atom MD simulations in explicit solvent with different force fields (OPLS [71], CHARMM [72], AMBER99SB [73], and GROMOS41a [74]) and water models (SPC [75] and TIP3P [76]). The sets of 100 models were subjected to the following procedure: (1) reconstruction and optimization using Modeller (see section 3.3), (2) short MD simulations in explicit solvent (simulation setup details are given at the end of this subsection), (3) binding energy calculation. The results are summarized in Table 3 (in terms of the lowest ligand-RMSD found in 10 top scored models) and compared to CABS-dock scoring. In general, all-atom force-fields provided better scoring results than the CABS-dock scoring scheme (except for the GROMOS43a1 force-field that performed better in two cases, comparably in one case and worse in two cases).

**Table 3. Lowest ligand-RMSD values found in 10 top scored models out of 100.** Several scoring schemes are compared (according to the CABS-dock server and interaction-energy values from all-atom MD with explicit solvent in different force-fields: CHARMM, AMBER99SB, GROMOS and OPLS) with the lowest ligand-RMSD values from the scored set of 100 (second column). The best scoring results among the different scoring methods are marked in bold.

| PDB code of a protein in unbound form used as an input | Lowest ligand-RMSD in 100 models subjected to scoring [24] | Lowest ligand-RMSD in 10 top-scored models out of 100 (in the brackets, model position in the scoring ranking is given) | | | | |
|---|---|---|---|---|---|---|
| PDB code | CABS-dock | CABS-dock | CHARMM | AMBER99SB | GROMOS | OPLS |
| 1CZZ | 2.58 | 3.21 (3) | **3.10** (9) | **3.10** (9) | 24.19 (1) | **3.10** (5) |
| 1N83 | 1.13 | 2.39 (3) | 2.29 (7) | 1.56 (1) | **1.13** (1) | **1.13** (5) |
| 2AM9 | 1.69 | 2.14 (4) | 2.03 (8) | **1.73** (6) | 2.16 (9) | 2.04 (7) |
| 2H14 | 3.09 | 4.70 (4) | **3.09** (3) | **3.09** (1) | 3.75 (1) | 3.75 (2) |
| 2HWQ | 1.36 | **1.49** (1) | 1.88 (9) | 1.88 (7) | 1.88 (8) | 1.88 (5) |

Additionally, for the 1N83 modeling case, in Figure 10 we present ligand-RMSD as a function of binding energy according to the coarse-grained CABS force-field (Figure 10a) and different all-atom force-fields (Figure 10b) (note that the final CABS-dock ranking is essentially based on structural clustering, not energetic scoring). As shown in the figure, the coarse-grained force-field provides a similar ranking of different quality models as all-atom force-fields in terms of general tendencies, i.e. a few of near-native models (having ligand-RMSD values lower than 5 Å) are on average ranked better than the other models of lower quality. However, all-atom force-fields are more efficient in discriminating the lowest ligand-RMSD models from the near-native models. As presented in Table 3, in the case of GROMOS and OPLS, the best quality model was ranked among the 10 top-scored models.

**Figure 10. Scoring results for the 1N83 case using interaction energy values.** (a) ligand-RMSD versus CABS energy plot for the 1N83 system. (b) ligand-RMSD versus all atom energies plots for the 1N83 system. Red, green, blue and magenta points correspond to AMBER99SB, GROMOS43a1, OPLS and CHARMM force fields, respectively.

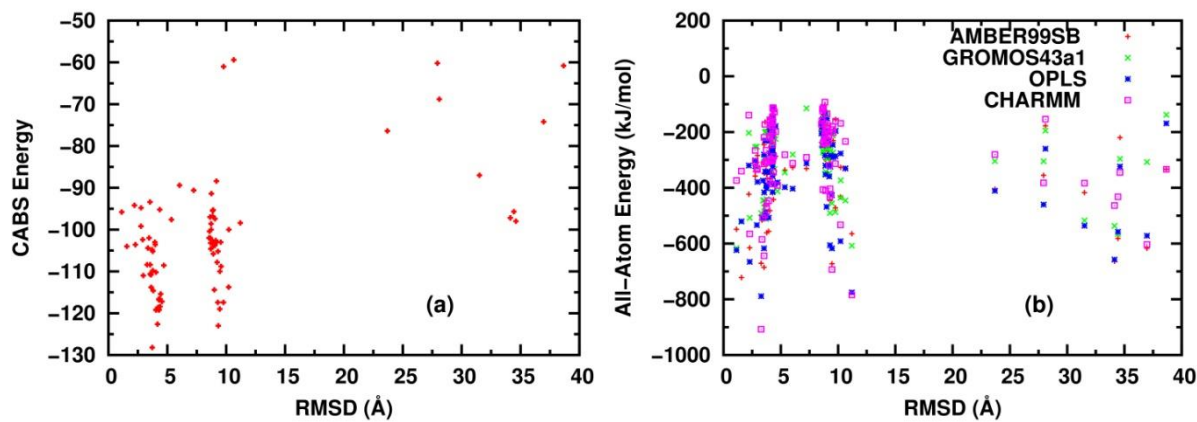

Although the above scoring results are promising, larger studies will be required using entire benchmark results and/or larger sets of CABS-dock generated models: 1000 or 10,000 (in which the best quality models were not identified using CABS-dock scoring [24]).

### 3.4.2 Details of the all-atom MD scoring procedure

In the described scoring exercises, we used four popular force fields for all-atom modeling: OPLS [71], CHARMM [72], AMBER99SB [73], and GROMOS41a [74] in conjunction with water models: SPC (Simple Point Charge) [75] (for GROMOS43a1, AMBER99SB and OPLS) and TIP3P [76] (for CHARMM). As all-atom MD modeling results might be force-field and water-model dependent [77, 78], it is always advisable to check different force-field schemes.

Simulations were carried out using a double precision version of the Gromacs-4.6 package [79]. The simulation details are as follows. The receptor-peptide system was placed in a dodecahedral box of a size that the minimal distance between the system and any periodic box edge was 1.0 nm, followed by filling up the box with water molecules. An appropriate number of ions were added to neutralize system charge. To avoid improper structures, the whole system was minimized with the steepest-descent method, before being equilibrated at 300K with two successive molecular dynamics runs, 10 ps each; the first one at constant volume (NVT equilibration), the second at constant pressure (NPT equilibration). The positions of the receptor and ligand were restrained by a biasing potential during NVT and NPT equilibrations for restrained moves of the receptor and ligand and unrestrained moves of water molecules. The pre-equilibrated conformations were used as starting structures for 20 ps production MD simulations. During a production run the ligand restraints were released, while the receptor was kept restrained. Temperature (300K) and pressure (1atm) were controlled by a v-rescale thermostat [80] and a Berendsen barostat [81], respectively.

We used periodic boundary conditions and calculated electrostatic interactions by means of the Ewald particle mesh method [82]. Non-bonded interaction pair lists are updated every 10 fs, using a cutoff of 1.4 nm. All bond lengths are constrained with the linear constraint solver LINCS [83] enabling the integration of the equations of motion with a time step of 2 fs. Initial velocities of the atoms were generated from Maxwell distribution at 300K. Data analysis was done using corresponding Gromacs tools.

All-atom interaction energy was calculated as a sum of Coulomb and Lennard-Jones interaction energies between the receptor and ligand and averaged over all saved configurations of a simulation run. During the simulation run atomic coordinates were saved every 0.2 ps. We obtained an average speed of ~0.5ns/day/CPU (Intel(R) Xeon(R) E5649 2.53GHz). Approximately 1 CPU hour is required for a single model 20 ps simulation run (for multiple CPUs, a speed of up to 5-6 can be achieved).

## 4. SUMMARY

In this paper, we demonstrated the performance and functionalities of the CABS-dock method and web server (server details are described in [24]). The unique CABS-dock capabilities of efficient docking of fully flexible peptides to flexible proteins without prior knowledge of the binding site make it a promising approach for modeling protein-peptide interactions. The promising extensions of the CABS-dock method include:

1. Combination of CABS-dock as a tool for initial peptide pose generation with the methods for high resolution refinement of protein-peptide interactions (e.g. HADDOCK [19] or Rosetta FlexPepDock [17]).
2. Extension of the CABS-dock procedure by exact scoring methods (as discussed in section 3.4) for better selection of accurate models out of large sets of generated models.
3. Incorporation of experimental data into the modeling process (such as NMR shifts, amide proton exchange [84, 85], phage display peptide libraries and mutational alanine scanning [86], or combination with other bioinformatics approaches enabling identification of binding site or any protein-peptide interaction features).
4. Increasing the flexibility of appropriate receptor fragments (not user defined as it is possible now, but automatic using for example the CABS-flex approach [37, 39] or bioinformatics-based identification of flexible receptor fragments).

The potential applications of CABS-dock embrace a vast array of strategies for the characterization of protein-peptide but also protein-protein interactions (PPIs), important for modern drug design. CABS-dock is a method for global and cost-effective search for interaction 'hot spots' or 'hot segments' [87], or binding areas that dominate the interaction. It is also possible to search for specific interaction sites by applying a fragment of the protein or peptide of known binding function as a query. Such an approach could be applied for peptide design that could specifically block PPIs with proven involvement in human disease and thus improve drug development [88, 89]. Importantly, many proteins engaged in PPIs are considered "undruggable" [90]. Such status has been assigned due to the lack of an internal cavity in these protein structures (that could be occupied by a small organic molecule), however they interact through flat and expanded binding sites, and thus may be targeted by peptide/protein therapeutics. CABS-dock prediction capabilities (and particularly the advanced option of excluding user-selected binding modes from docking search) can also be useful in computational epitope mapping, where multiple binding sites have to be considered [91]. It must be emphasized that the development of epitope-based vaccines and other antigen-based drugs is not possible without clear identification of the antigenic region involved in binding [92]. In vivo epitope mapping can be challenging due to difficulties in expressing and purifying antigens such as membrane proteins and complexes [93]; thus, progress in reliable computational methods is greatly anticipated.

# APPENDIX

The authors acknowledge Dr Michal Jamroz for his valuable support in programming modeling scripts, crafting online documentation, constant improving of the CABS-dock systems and for cheerful discussions.

The Appendix contains a tutorial for CABS-dock results visualization and analysis in VMD [25] (a molecular visualization program for displaying, animating, and analyzing large biomolecular systems using 3-D graphics and built-in scripting).

# ACKNOWLEDGMENTS

This work was supported by the Foundation for Polish Science TEAM project (TEAM/2011-7/6) co-financed by the European Regional Development Fund operated within the Innovative Economy Operational Program; National Science Center grant [MAESTRO 2014/14/A/ST6/00088]; Polish Ministry of Science and Higher Education Grant No. IP2012 016872.

**APPENDIX - Tutorial for CABS-dock results visualization and analysis**

*Appendix Table Of Contents*

## 1. Introduction

This tutorial describes how to:

- load all the necessary files into the VMD molecular graphics program
- create simple graphic representations for viewing trajectories
- align models from trajectories with a known receptor structure
- calculate peptide RMSDs (root mean square deviation from the reference structure)
- create plots from calculated RMSDs vs. energies from the CABS-dock results

The following software is required:

- <u>VMD</u> with an RMSD Trajectory Tool plug-in (included in VMD ver. 1.8.8 and higher). VMD is user friendly yet advanced software for the analysis and visualization of structure and molecular dynamics trajectories of biological systems. In this tutorial some features are omitted, e.g. VMD for basic protein analysis: Sequence Viewer, Contact Map or Ramachandran Plot. Those plug-ins are very well documented on the <u>VMD website.</u>
- <u>gnuplot</u>· gnuplot is software for creating high quality plots from user data.

Both programs are free to use, available for major operating systems and fairly easy to install.

## 2. CABS-dock result files and reference structure

In this tutorial, we used an example prediction run described in the manuscript (in section 3.1.1). The example was created using the following input data:

- **Peptide sequence:** SSRFESLFAG
- **Peptide secondary structure:** CHHHHHHHHC
- **Receptor input structure:** PDB ID: 2AM9, crystal structure of human androgen receptor in the unbound form (without a peptide)

Reference structure used for the calculation of RMSD values to the experimentally determined peptide-bound structure:

- **Peptide-receptor complex structure:** PDB ID: 1T7R, crystal structure of human androgen receptor in complex with the peptide

Typically the structure of the bound complex is unknown, so one of the resulting models may be used as a reference (for example the top scored model: model_1.pdb).

The results page of this job is available at:
http://biocomp.chem.uw.edu.pl/CABSdock/job/7f0bda72050182.

First, from the "Docking predictions results" tab, download a compressed archive with the output data for further analysis. The archive contains the following files:

- `model_*.pdb` – 10 final models, numbered from 1 to 10 (in the PDB file format and all-atom representation)
- `cluster_*.pdb` – cluster models (groups of models that have been classified in structural clustering to particular clusters), numbered from 1 to 10 (in the PDB file format and C-alpha representation). Cluster numbering corresponds to models numbering, i.e. model_7.pdb is a representative model for models grouped in the seventh cluster (ranked as seventh) (cluster_7.pdb).
- `trajectory_*.pdb` – complete set of 10 trajectories, numbered from 1 to 10 (in the PDB file format and C-alpha representation). Each trajectory contains 1000 models.
- `top1000.pdb` – top 1000 models (selected for further clustering and analysis) from the 10 trajectories (in the PDB file format and C-alpha representation)
- `input.pdb` – input structure of the receptor
- `README` – log file with all information to recreate the simulation
- `energy.txt` – log file with energy values (from the coarse-grained CABS force-field) for all resulting models; columns contain the following data (in the order listed):
  - trajectory (replica) index (`no`)
  - trajectory (replica) frame index (`fr`)
  - temperature - a parameter in the CABS model that controls the acceptance of new conformations within the Monte Carlo method (`temp`)
  - energy value for the protein receptor only (`recE`)
  - energy value for the peptide only (`pepE`)
  - energy value for the protein-peptide interaction only (`rec-pepE`)

- energy value for the entire protein-peptide complex (`totalE`)

The energy.txt file contains the following data (column headers are described above):

```
      no    fr    temp    recE       pepE     rec-pepE   totalE
       1     1    1.95  -2261.77    -37.87      0.00    -2299.64
       1     2    1.95  -2255.36    -38.43      0.00    -2293.79
       1     3    2.45  -2195.35    -21.44      0.00    -2216.79
       1     4    2.45  -2193.94      0.49   - 17.80    -2211.25
       1     5    2.95  -2160.39    -36.25     -5.60    -2202.24
                              ...
      10   996    4.00  -1920.98    -14.86    -32.10    -1967.94
      10   997    4.50  -1965.89     -3.03    -30.00    -1998.92
      10   998    4.00  -1959.18     -5.52    -42.40    -2007.10
      10   999    4.00  -1995.16    -14.23    -32.20    -2041.59
      10  1000    4.00  -1961.22    -14.34    -43.60    -2019.16
```

Finally, download the 1T7R structure from the RCSB PDB database.

### 3. Using VMD

**Below, different font styles are used to mark different features:**
- *Italics: buttons, menu positions, fields, lists, sections*
- `Lucida Console font: CONSOLE INPUT, VALUES, FILES`
- Underlined: VMD window panels

**Loading files**
Start VMD. Three application windows should be displayed: console, OpenGL Display and VMD Main. Load necessary files using one of the following methods: using the graphical interface (see section 0) or via the console (see section 0).

**From the graphical interface**

Click *File>New Molecule...*. From the Molecule File Browser window > select *Browse*. In the Choose a molecule file window navigate to server results, select file `trajectory_1.pdb` and click *Open* (or *OK*).

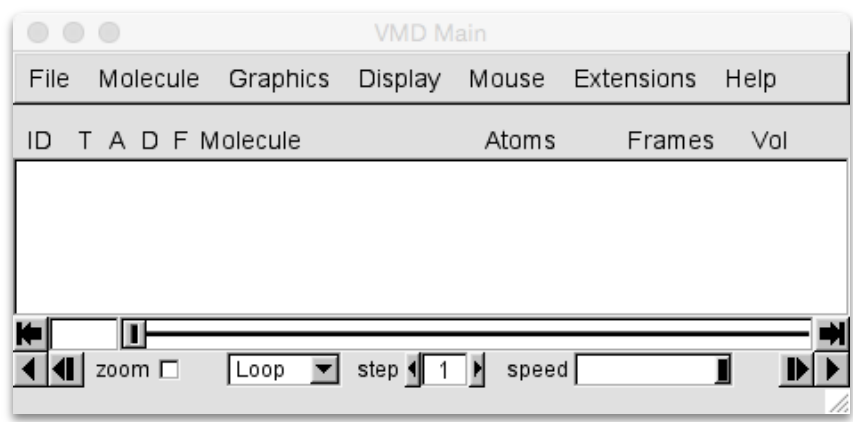

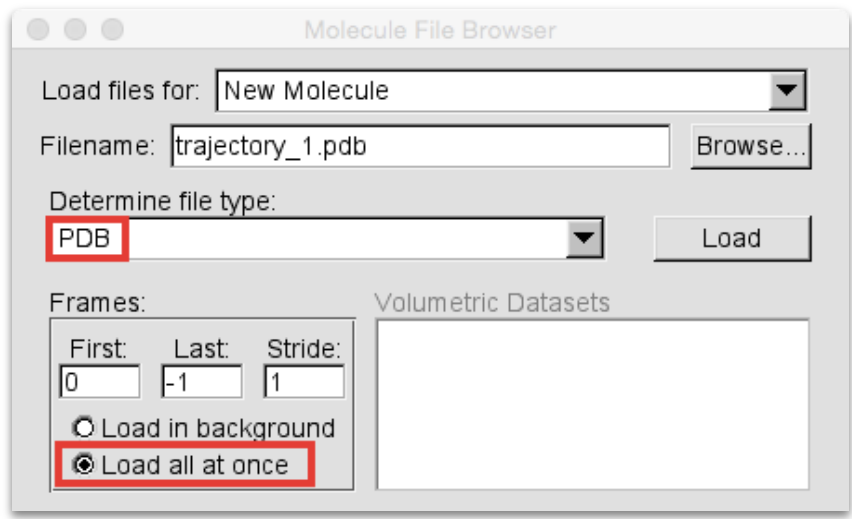

In the Molecule File Browser in the *Determine file type* field select `PDB`. Choose *Load all at once* in the *Frames* section and click *Load*.

To load the rest of trajectories select `trajectory_1.pdb` in the VMD Main window and choose *File>Load Data Into Molecule* and repeat previous steps in the Molecular File Browser window.

**It is important to load trajectories in the following order:** `trajectory_2.pdb`, `trajectory_3.pdb`, ..., `trajectory_10.pdb` **for further analysis (plotting RMSDs vs. Energy).**

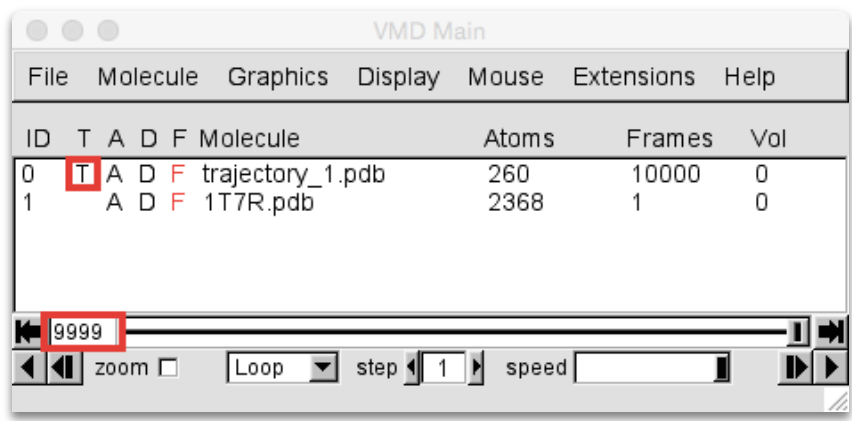

The VMD Main window frame indicator should show: 9999 due to VMD numbering from 0 to 9999 for the total of 10000 frames. Load the reference structure (`1T7R.pdb`) as a new molecule. Set trajectories "top"- in the VMD Main window click under column marked "T" in the trajectories line. The "top" state is marked with "T"

**Via the console**
An alternative (and faster) way of loading all files is through *Tk Console*.

From the VMD Main window click *Extensions>Tk Console*. In the VMD TkConsole window navigate, using the `cd` command (change directory – e.g.

`cd /home/user/Documents/results/`) to a folder with server results.

Then execute following commands:

```
for { set i 1 } { $i <= 10 } { incr i } {mol addfile trajectory_$i.pdb wait for all }
mol new 1T7R.pdb
mol top
```

```
>Main< (CABSdock_7f0bda72050182) 3 % for { set i 1 } { $i <= 10 } {incr i }
{ mol addfile trajectory_$i.pdb waitfor all }
>Main< (CABSdock_7f0bda72050182) 4 % mol new 1T7R.pdb
1
>Main< (CABSdock_7f0bda72050182) 5 % mol top 0
>Main< (CABSdock_7f0bda72050182) 6 %
```

## Changing graphical representations

This section describes how to change the default representation (see screenshot on the right) to a more convenient one.

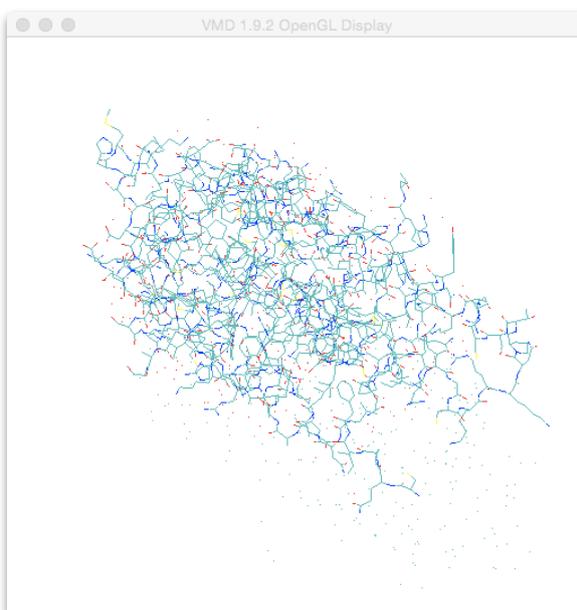

## Chain identification

First, it is necessary to find chain identifiers.

From the VMD Main window *Graphics*>select *Representations….*

In the Graphical Representations window choose *Selections* tab >*Keyword* list >*chain*.

The characters that appear in the Values section are chain IDs of the molecules in the currently selected molecule (*Selected Molecule* list).

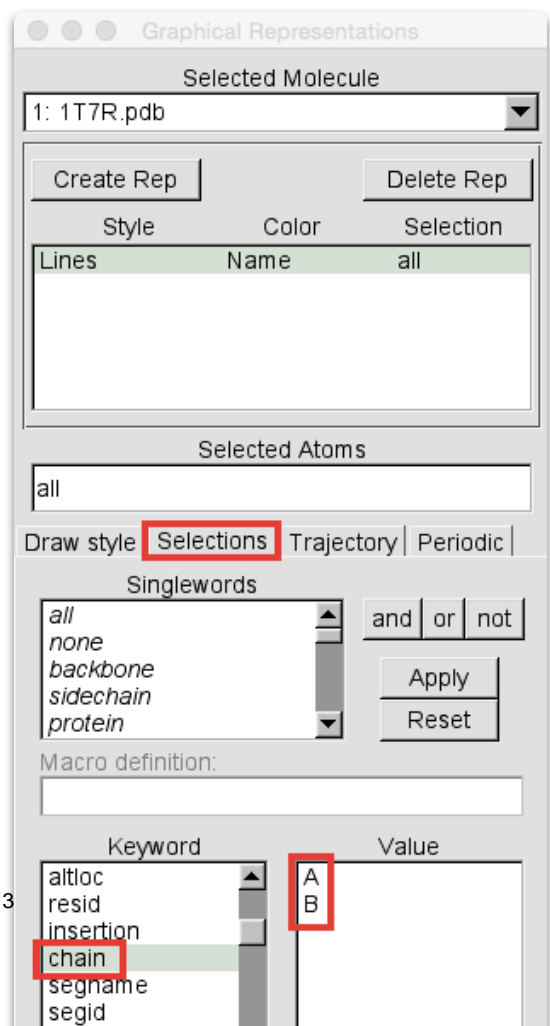

**It is crucial to remember or note chain IDs for further analysis (molecule fragment selection).**

Find the chain IDs for the second molecule. In the example, both the trajectories and the reference contain chains A and B. A more complicated system may contain more chains in the receptor molecule but the peptide always has only one (usually the shortest) chain.

### Creating trace representation

Now delete the existing representations by selecting entries on the list and click *Delete Rep*. In the *Selected Atoms* field select `chain A` and click *Create Rep*. For every representation set *Drawing Method* on *Trace*, *Coloring Method* on *ColorID* and choose color. Repeat the procedure for the rest of the chains and all other molecules from the *Selected Molecule* list. Now it is possible to view all the models from trajectories in the OpenGL Display window using the VMD Main frame slider.

Colors in the example screenshot below: red - 1T7R receptor, blue - receptor from the trajectory, green - 1T7R peptide, orange – peptide from the trajectory.

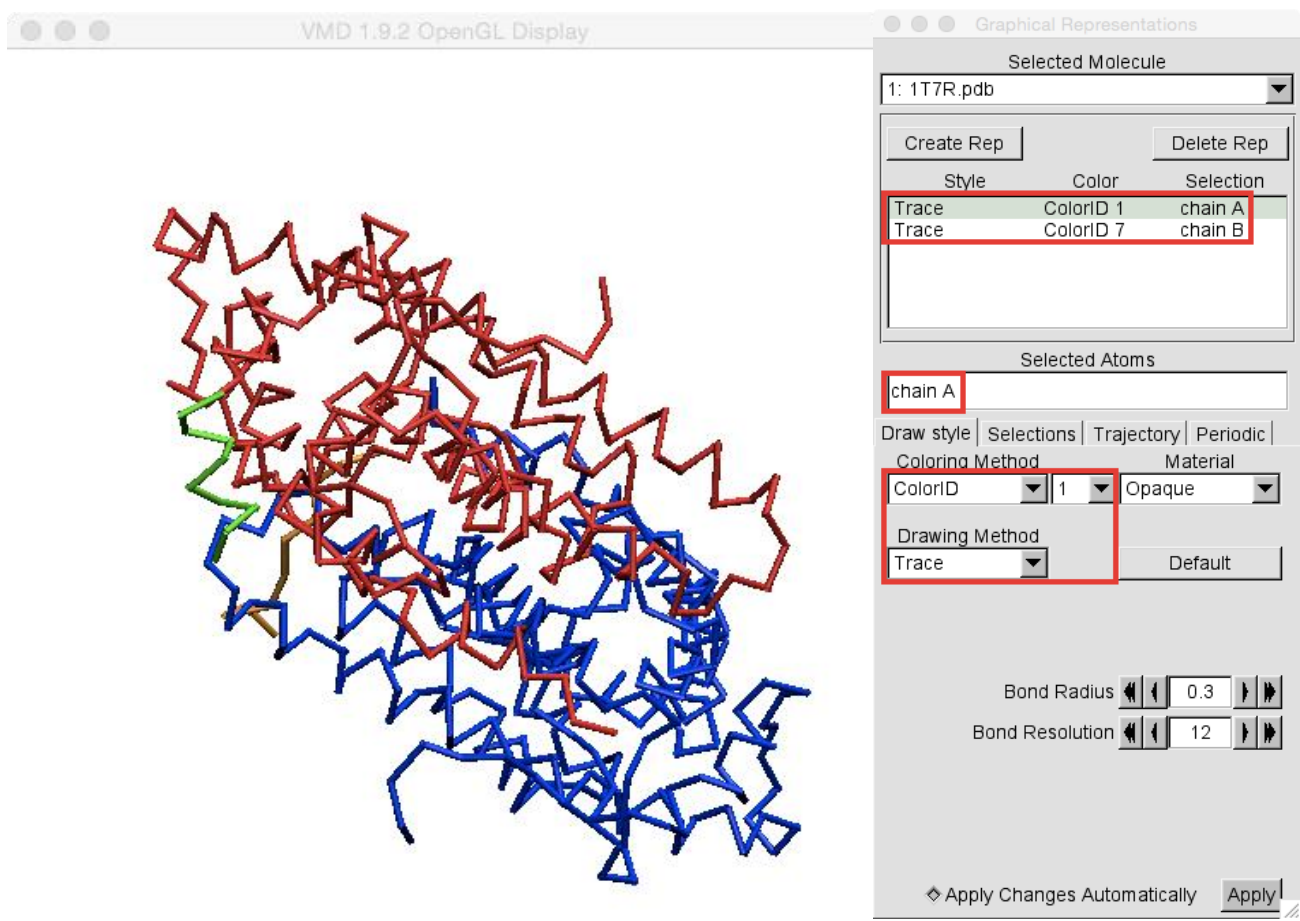

# RMSD analysis
**This part concerns calculating peptide-based RMSDs through 10,000 models of an example trajectory.**

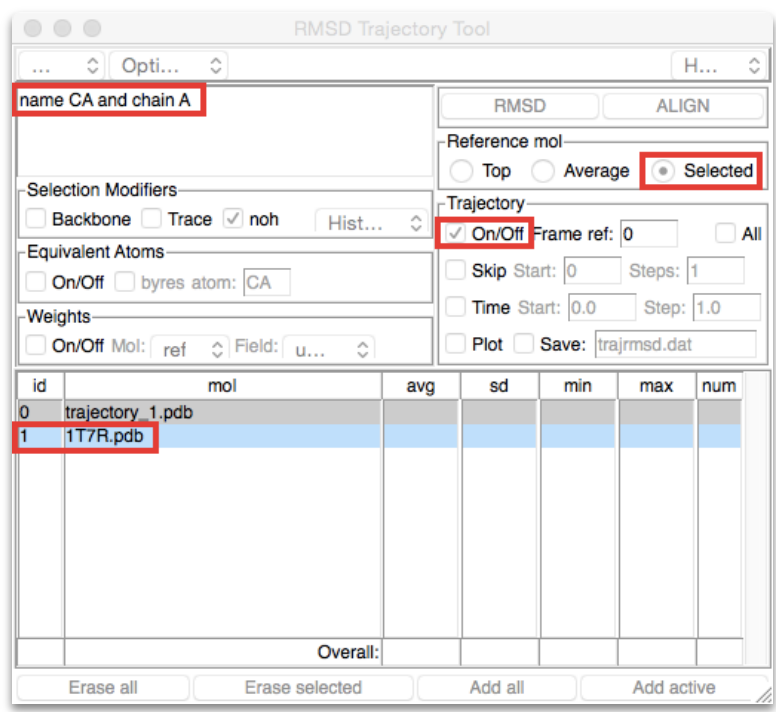

From the VMD Main window select *Extensions>Analysis>RMSD Trajectory Tool*. The RMSD Trajectory Tool window should contain names of two files: `trajectory_1.pdb` and `1T7R.pdb`. If not, click *Erase all* and *Add all*. In the *selection frame* (in the top left corner of the window) type: `name CA and chain A`. This command selects alpha carbon atoms from the protein backbone from chain A (receptor).

Set *Reference mol* to *Selected* and select the *Trajectory>On/Off* checkbox and click *Align*.

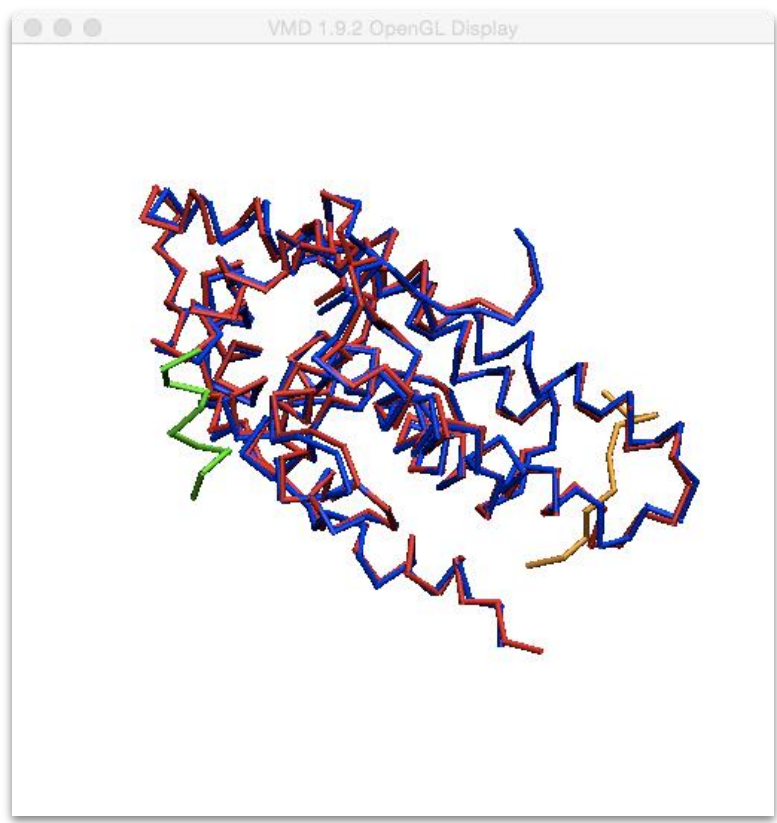

In the OpenGL Display window you will see that receptors from both systems are aligned.

Back in the RMSD Trajectory Tool window in the *selection frame* type: `name CA and chain B`. This selects alpha carbon atoms from the protein backbone from chain B

(peptide). Click *RMSD*.

Finally, from the RMSD Trajectory Tool window menu choose *Save data...* and save data as a text file: `trajrmsd.txt` in the results folder.

The file contains frame IDs and calculated RMSDs:

```
frame    mol0
    0   34.489
    1   33.945
    2   25.595
    3   27.096
    4   30.941
    ...
 9997   48.781
 9998   45.021
 9999   42.475
```

It is also possible to view and save a simple RMSD vs. Frame plot by choosing *Plot data*.

## Trajectory movie

After aligning receptor structures, it is easy to observe conformational sampling of the peptide.

In the OpenGL Display window click and hold the left mouse button to rotate representations to expose the peptide from the reference structure.

Buttons in the VMD Main window: *Play forward* and *Play in reverse* allow the user to view the whole trajectory as a movie. To adjust speed or frame skipping use the speed slider and step counter.

To find a particular frame use *Step forward*, *Step in reverse* or use the frame slider by clicking on it and holding the left mouse button to drag the frame marker.

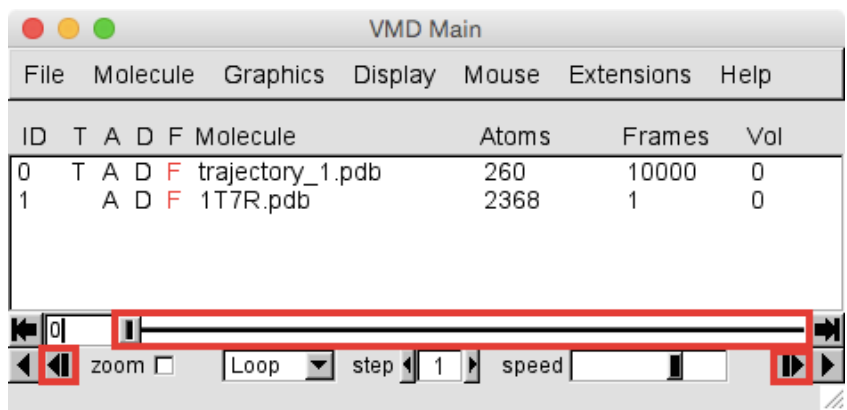

VMD also enables exporting the trajectory via the Movie Maker plugin: VMD Main>*Extensions*>*Movie Maker* but it requires an external encoder (e.g. ffmpeg or mencoder for Linux/OS X, VideoMach for Windows) and preferably an external renderer.

## 4. Creating plots

### File preparation

This part is about file preparation for gnuplot, for the Linux/OS X console. Windows users are encouraged to use text editors with block editing capabilities for merging files (e.g. Komodo Edit, Sublime Text 2 or Edit Plus, vim).

Before plotting with gnuplot the data from `energy.txt` and `trajrmsd.txt` have to be combined into one file. First remove the header from `trajrmsd.txt`:

`sed 1d trajrmsd.txt > tmprmsd.txt`

`paste energy.txt tmprmsd.txt > enermsd.txt`

The new file will resemble the sample (in the actual file there will be no header):

```
    no    fr temp     recE    pepE rec-pepE   totalE    VMD-fr    RMSD
     1     1 1.95 -2261.77  -37.87     0.00 -2299.64         0   46.085
     1     2 1.95 -2255.36  -38.43     0.00 -2293.79         1   43.458
     1     3 2.45 -2195.35  -21.44     0.00 -2216.79         2   43.565
     1     4 2.45 -2193.94    0.49   -17.80 -2211.25         3   33.113
     1     5 2.95 -2160.39  -36.25    -5.60 -2202.24         4   38.857
                              ...
    10   998 4.00 -1959.18   -5.52   -42.40 -2007.10      9997   22.368
    10   999 4.00 -1995.16  -14.23   -32.20 -2041.59      9998   17.720
    10  1000 4.00 -1961.22  -14.34   -43.60 -2019.16      9999   16.552
```

The extreme values for RMSD and energies can be found quickly using `sort` and `head` commands:

`sort -n -k9,9 enermsd.txt | head -n10` gives top 10 records with the lowest RMSD

`sort -n -k6,6 enermsd.txt | head -n10` gives top 10 records with the lowest interaction energy

The value in the `-k` parameter indicates which column is used for sorting.

### Using gnuplot

Start gnuplot (e.g. by typing `gnuplot` in the console in the results folder) and set a proper terminal size, e.g. jpeg with 800x640 plot resolution:

`gnuplot> set terminal jpeg size 800,640`

```
gnuplot> set output "rmsdene.jpg"
gnuplot> set xlabel "RMSD"
gnuplot> set ylabel "Interaction Energy"
gnuplot> set title "Energy vs RMSD"
gnuplot> plot "enermsd.txt" using 9:6 with dots notitle
```

This will produce a 800 by 640 JPEG `rmsdene.jpg` file with a dotted scatter plot titled "Energy vs. RMSD", its X axis labeled "RMSD" and Y axis labeled "Interaction Energy". Numerical values in the last command indicate the columns used in format "X:Y", so changing it to `8:9` will create a RMSD vs. frame plot and to `8:6` will create an Interaction energy vs. frame plot. Do not forget to also change the appropriate filename, title and labels. For other options please check the gnuplot documentation.

See the aforementioned figures below:

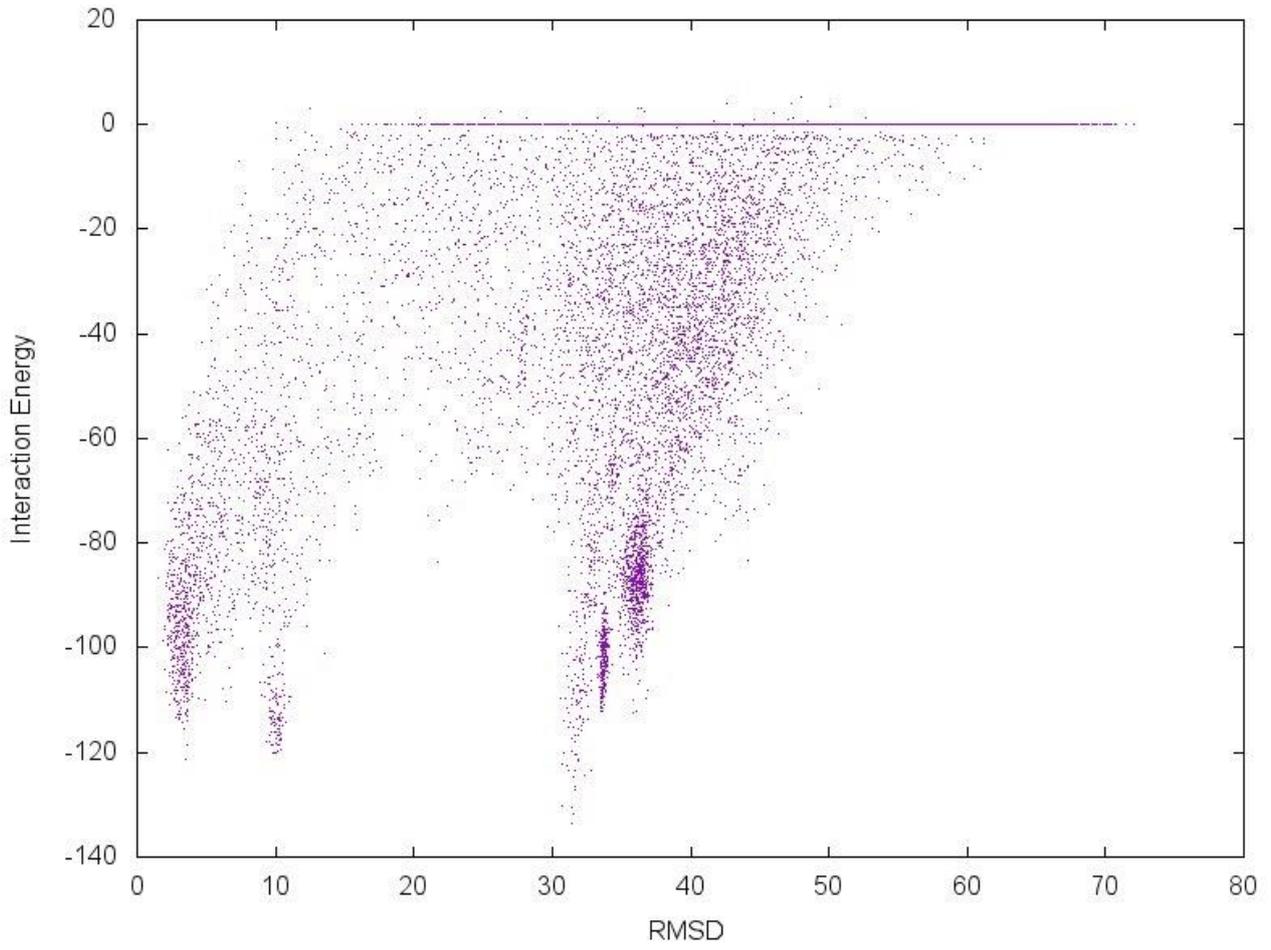

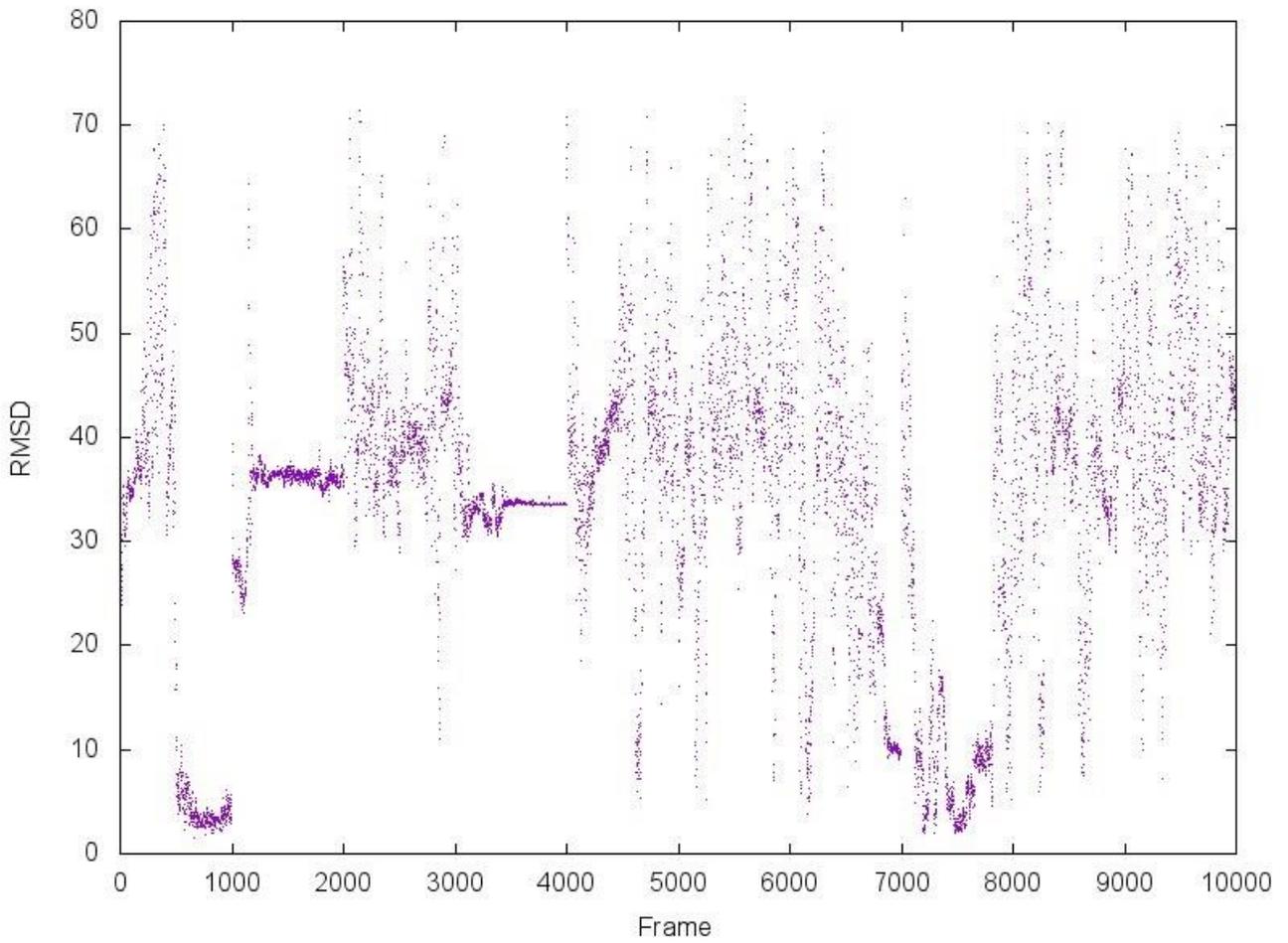

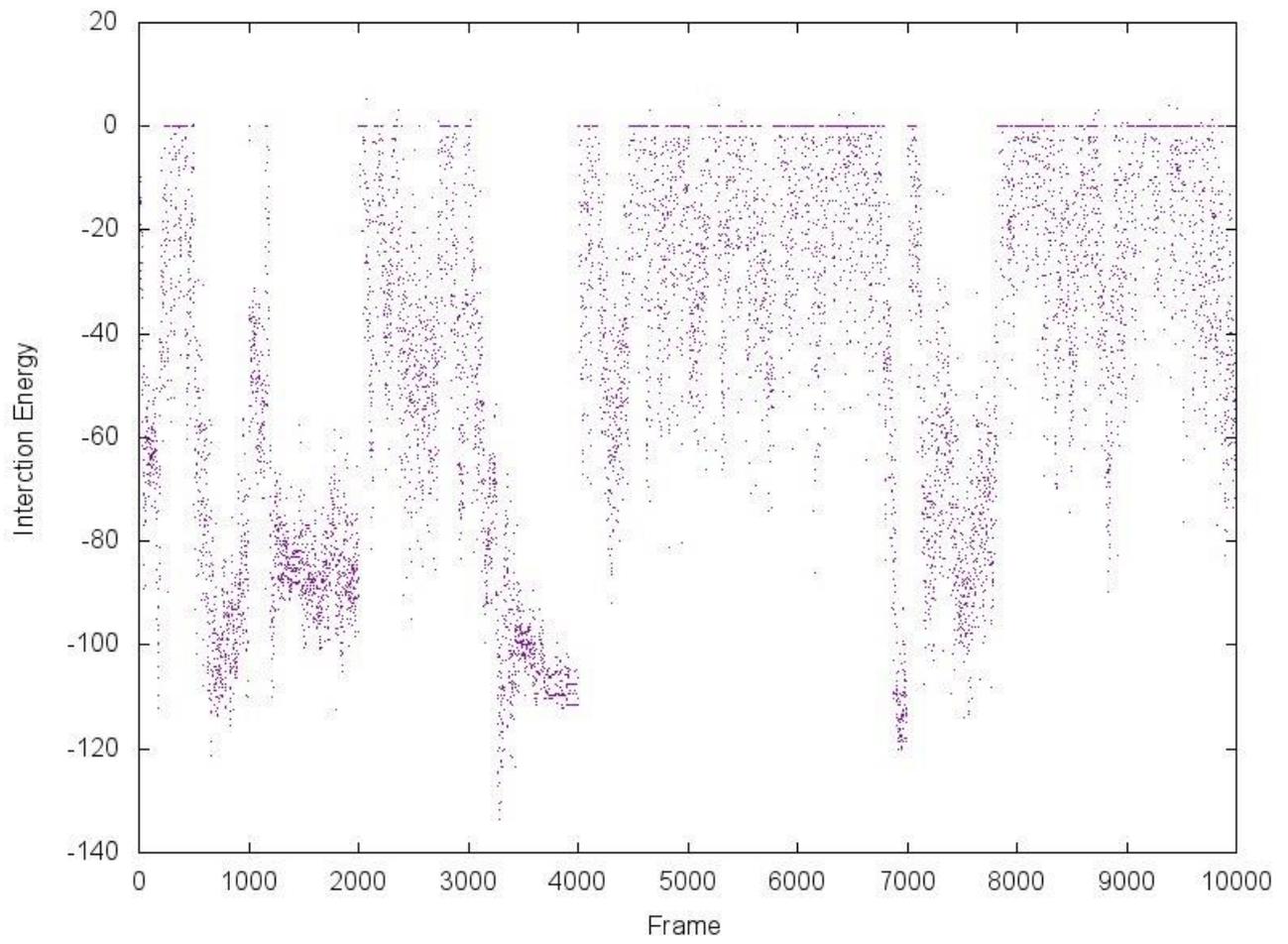